\documentclass{acm_proc_article-sp}
\usepackage{graphicx} 
\usepackage{amsmath} 
\usepackage{amssymb}
\usepackage{multirow}
\usepackage{subfigure}
\usepackage{hyperref}
\newcommand{\BEQ}{\begin{equation}}
\newcommand{\EEQ}{\end{equation}}
\newcommand{\BEA}{\begin{eqnarray}}
\newcommand{\EEA}{\end{eqnarray}}
\newcommand{\be}{\begin{eqnarray}}
\newcommand{\ee}{\end{eqnarray}}
\newcommand{\benn}{\begin{eqnarray*}}
\newcommand{\eenn}{\end{eqnarray*}}

\usepackage{color}

\begin{document}
\sloppy
\frenchspacing

\title{Information-Theoretic Measures of Influence \protect\\ Based on Content Dynamics}

\numberofauthors{2} 
\author{
\alignauthor
Greg Ver Steeg\\
       \affaddr{Information Sciences Institute}\\
       \affaddr{University of Southern California}\\
       \affaddr{Marina del Rey, California}\\
       \email{gregv@isi.edu}
\alignauthor
Aram Galstyan\\
       \affaddr{Information Sciences Institute}\\
       \affaddr{University of Southern California}\\
       \affaddr{Marina del Rey, California}\\
       \email{galstyan@isi.edu}\date{30 July 2012}
}

\maketitle
\begin{abstract}
The fundamental building block of social influence is for one person to elicit a response in another. 
Researchers measuring a ``response'' in social media typically depend either on detailed models of human behavior or on platform-specific cues such as re-tweets, hash tags, URLs, or mentions. 
Most content on social networks is difficult to model because the modes and motivation of human expression are diverse and incompletely understood. 
We introduce {\em content transfer}, an information-theoretic measure with a predictive interpretation that directly quantifies the strength of the effect of one user's content on another's in a model-free way. 
Estimating this measure is made possible by combining recent advances in non-parametric entropy estimation with increasingly sophisticated tools for content representation. 
We demonstrate on Twitter data collected for thousands of users that content transfer is able to capture non-trivial, predictive relationships 
even for pairs of users not linked in the follower or mention graph.
We suggest that this measure makes large quantities of previously under-utilized social media content accessible to rigorous statistical causal analysis.
\end{abstract}



\section{Introduction}

While the emergence of various online social networking platforms provides a steady source of data for researchers, it also provides a source of constantly evolving complexity. 
Most prior research has focused on analyzing various static topological properties of networks induced by social communication, while discarding the content of communication. At the same time, there is a growing recognition that a more nuanced understanding of social interactions requires analyzing the semantic content of communications.  
For instance,  it has been suggested that  linguistic cues in communicative patterns, as well as the ways individuals echo and accommodate each other's linguistic styles,  can be indicative of relative social status of participants~\cite{Danescu-Niculescu-Mizil2012WWW}. 
Despite recent progress, however, content-based analysis of social interactions is still a challenging problem due to the lack of adequate quantitative methods for extracting useful signals from unstructured text. Another significant hurdle  is that the design and usage of social networks, and thus the interpretation of various signals, are changing over time.

Here we suggest a novel, information-theoretic approach for leveraging user-generated content to characterize  interactions among social media participants. Specifically, given all the content generated by a set of users (e.g., a sequence of tweets), our goal is to find meaningful edges that indicate social interactions among this set of users. Our approach is model-free in the sense that it does not presuppose a particular behavioral model of users and their interactions. Instead, we view users as producers of some arbitrarily encoded information stream. If $Y$'s stream has an effect on $X$'s, then access to $Y$'s signal can, in principle, improve our prediction of $X$'s future activity. This is what we mean by a {\em predictive link}.  We show that this general notion of predictability can be used to identify social influence.

\begin{figure}[htbp] 
   \centering
   \includegraphics[width=0.8\columnwidth]{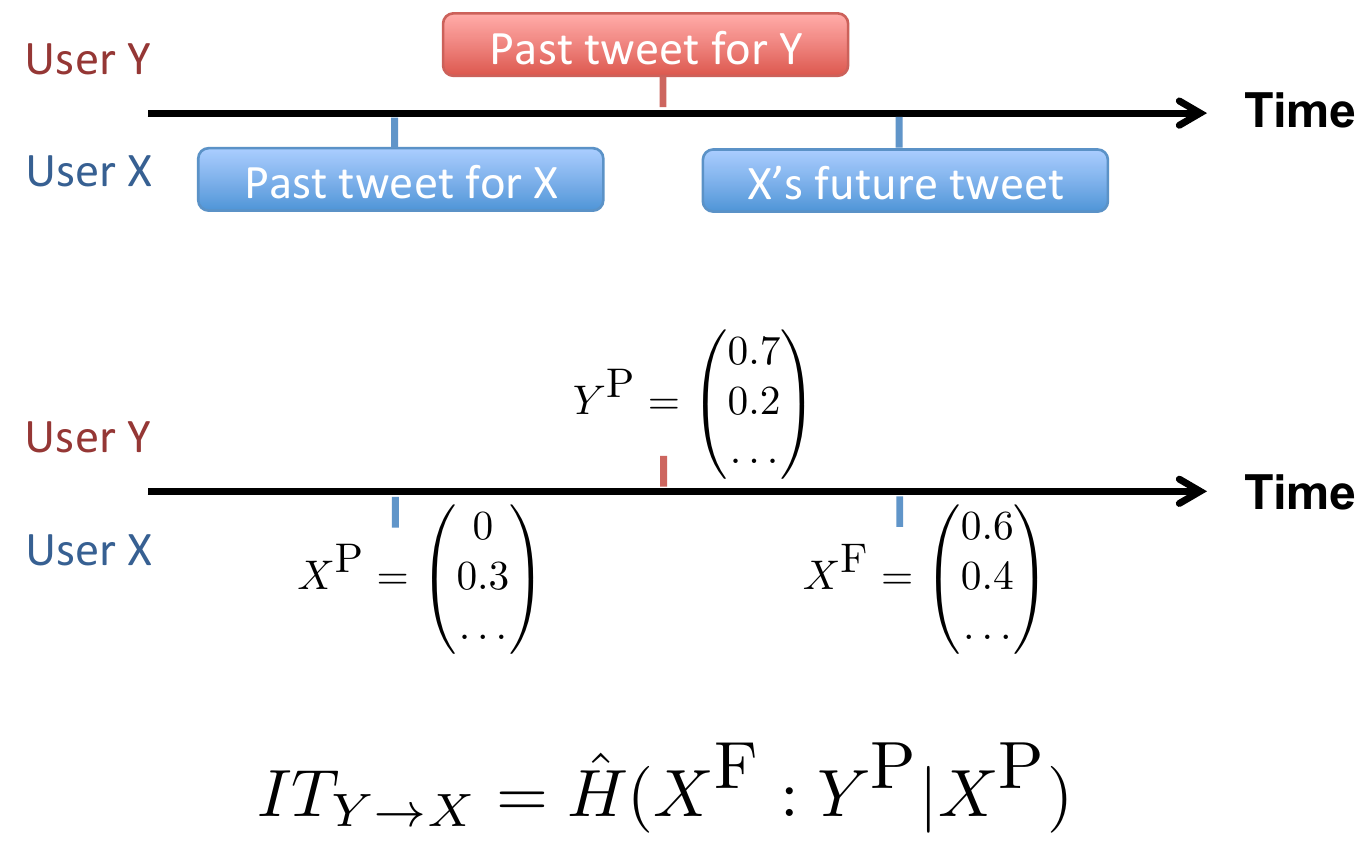} 
   \caption{Is the content of X's future tweet, $X^F$, predictable from past tweets, $Y^P,X^P$? We answer this question by first transforming the text of tweets into vectors. Joint samples of these variables can be used to estimate information transfer, or transfer entropy, quantifying how predictive $Y$'s tweets are for $X$'s future tweets.}
   \label{fig:schematic}
\end{figure}

The technical approach proposed here consists of two main ingredients (see Fig.~\ref{fig:schematic}). First, we represent user-generated content in a high-dimensional space so that a sequence of user-generated posts is mapped to a time-series in this space. Second, we apply information-theoretic measures to those time series to discover and quantify directed influence among the users.    
Because our method is based on information-theoretic principles, it is easy to interpret, applicable to arbitrary signals and/or platforms, and flexible with respect to the representation of content.

Our approach ultimately reduces to calculating an information-theoretic measure called {\em transfer entropy} between pairs of stochastic processes~\cite{Schreiber2000}. Intuitively, transfer entropy between processes $X$ and $Y$ quantifies how much better we are able to predict the target process $X$ if we use the history of the process $Y$ and $X$ rather than the history of $X$ alone. By using transfer entropy as a statistical measure of the relationship between the content of Y's tweets and the content of X's subsequent tweets, we construct a graph of {\em predictive links}, based only on the content of users' tweets. Our results demonstrate that transfer entropy indeed reveals a variety of predictive, causal behaviors. Surprisingly, we also discover that many of the most predictive links {\em are not present in the social network}, through mentions nor friend links. 
Nevertheless, in Sec.~\ref{sec:mention}, we verify the meaningfulness of our measure by showing that predictive links are a statistically significant predictor of mentions on Twitter.  

To summarize, our main contribution is a novel application of an information-theoretic framework to content-based social network analysis, providing a general, flexible measure of meaningful relationships in the network. This construction is made possible by two apparently novel technical insights. (1) Current state-of-the-art methods for estimating entropic measures such as mutual information continue to perform well in high-dimensional spaces as long as they are effectively low-dimensional in some sense. (2) While content representations of user activity are high-dimensional, they are effectively low dimensional in the required sense. Taken together, these two points allow us to successfully apply entropic estimators in a previously inaccessible regime.

After motivating our technical approach and defining the relevant information-theoretic quantities in Section~\ref{sec:theory}, we describe how to estimate those quantities in Sec.~\ref{sec:TEestimation}, and demonstrate their use on real-world data from Twitter in Sec.~\ref{sec:results}. Finally, we  give an overview of related work in Sec.~\ref{sec:related}, followed by a discussion of results in Sec.~\ref{sec:conclusion}.

\section{Technical Approach}
\label{sec:theory}
\subsection{Motivation}
Let us consider a set of users that generate a time-stamped sequence of  text documents, e.g., tweets. Let X and Y be two such users. With a slight misuse of notation, let $X^P$ and $Y^P$ denote  the content of tweets generated by those users up to some time, in some representation. While our approach is not limited to a particular representation, below we will  use $X^P$ and $Y^P$ to describe  topical representation of the content (e.g., obtained via  Latent Dirichlet Allocation, or LDA).   

Consider now the problem of predicting the content of the next tweet generated by X, denoted by $X^F$; see Fig.~\ref{fig:schematic}. Generally speaking, $X^F$ is a random variable that can depend on a large number of factors that might  not be directly observable: topical interests of user X (and her friends), exogenous events, and so on.  Here, however, we are interested in the extent to which $X^F$ is  influenced by the past tweets $X^P$ and $Y^P$. Namely, we would like to see how much knowing the past content generated by user Y,  $Y^P$,  helps us to better predict $X^F$. If knowing Y's past tweets helps us to predict $X^F$ more accurately, then we can say that Y exerts certain influence on X. 

The notion of  {\em influence} (or {\em causality}) described above is taken in the sense of Granger causality~\cite{Granger1980} which demands that (1) the cause occurs before the effect; (2) the cause contains information about the effect that is unique, and is in no other variable~\cite{causalityreview}. In practice, determining that information is ``in no other variable'' is difficult. For determining a causal effect on a user in a social network, we only attempt to rule out the user's recent past as an explanation. Exogenous and long-term effects are difficult to account for but will be discussed in some interesting cases. The principle behind Granger causality was originally applied in the context of regression models, but applying these ideas in the context of information theory leads to effective tests of causality~\cite{causalityreview}.

\subsection{Transfer Entropy}\label{sec:preliminaries}

We denote by $H(X)$ the entropy of a random variable, $X$, with some associated probability distribution, $p(\vec x) \equiv \Pr (X=\vec x),$ for $\vec x \in \mathbb R^{d_x}$. In this case (differential) entropy is defined in the standard way, using the natural log,
$$ H(X) = \mathbb E (-\log p(x)) = -\int dx~ p(x) \log p(x). $$
We sometimes speak of entropy as quantifying our ``uncertainty'' about $X$.
Standard higher order entropies such as mutual information and conditional entropy can be defined in terms of differential entropy as $H(X:Y) = H(X)+H(Y)-H(X,Y)$ and $H(X|Y) = H(X,Y) - H(Y)$, respectively. Conditional information can be interpreted as the reduction of uncertainty in $X$ from knowing $Y$.

Transfer entropy, or information transfer~\cite{Schreiber2000}, can be defined as,
\begin{eqnarray}\label{eq:te}
IT_{Y\rightarrow X} &=& H (X^{ \mbox{\tiny F}}:Y^{ \mbox{\tiny P}}|X^{ \mbox{\tiny P}}) \\
&=&  H (X^{ \mbox{\tiny F}}|X^{ \mbox{\tiny P}}) - H (X^{ \mbox{\tiny F}}|Y^{ \mbox{\tiny P}},X^{ \mbox{\tiny P}}), \nonumber
\end{eqnarray}
where $X^{ \mbox{\tiny F}}$ is interpreted as information about user $X$'s future behavior, and $X^{ \mbox{\tiny P}},Y^{ \mbox{\tiny P}}$ as user $X$ and $Y$'s past behavior, respectively. The temporal indices dictate that cause should come before effect, and conditioning on $X$'s past insures that any explanatory value from $Y$ is not already present in $X$'s past behavior. The first line writes this quantity succinctly as a {\em conditional mutual information} while the second line has the nice interpretation that we are interested in how much knowing $Y^{ \mbox{\tiny P}}$ reduces our uncertainty about $X^{ \mbox{\tiny F}}$. This quantity is asymmetric, so in principle $IT_{Y\rightarrow X} \neq IT_{X\rightarrow Y}$. We will see examples where this is the case. 

In this paper, we take $Y^{ \mbox{\tiny P}},X^{ \mbox{\tiny P}},X^{ \mbox{\tiny F}}$ to be random processes representing the content of tweets for users $X$ and $Y$, and so we refer to this measure as {\em content transfer}. In particular, referring to Fig.~\ref{fig:schematic}, given some concrete procedure to turn an individual tweet into a vector, $\vec x$, we consider samples, $i=1\ldots N$, of triples of tweets $(\vec x^{ \mbox{\tiny F},i},\vec y^{ \mbox{\tiny P},i},\vec x^{ \mbox{\tiny P},i})$ representing $X$'s tweet, $Y$'s most recent previous tweet, and $X$'s most recent previous tweet. Note that we demand that $Y$'s tweet should occur after $X$'s previous tweet, otherwise the causal effect of $Y$'s tweet is already being taken into account as affecting $X$'s previous tweet . Also, $Y$ could tweet many times in between $X$'s tweets, but we only consider the most recent tweet for simplicity. 
Many possibilities exist to represent a tweet as a vector and in Sec.~\ref{sec:topic} we describe some of them along with the topic model approach used in this paper.

Note that we could drop the conditioning on $X^P$ to get the mutual information between $X$'s future and $Y$'s past, sometimes called time-delayed or time-shifted mutual information. We will compare this simpler quantity to transfer entropy below. On the other hand, in principle, we could add even more conditioning on other variables like news stories, other users' activity, $Z^{ \mbox{\tiny P}}$, or more history for users $X$ and $Y$.
The difficulty is in estimating these entropies from sparse data, and adding more conditions also increases the dimensionality of the problem. 

\section{Estimating Transfer Entropy}
\label{sec:TEestimation}
Generally speaking,  calculating entropic measures for high-dimensional random variables is problematic due to data sparsity~\cite{panzeri}. Rather than binning data and estimating probability distributions as a prerequisite for calculating entropy,  Kozachenko and Leonenko~\cite{Kozachenko} introduced an entropy estimator that was asymptotically unbiased and did not require binning of data. Binless estimators were extended to higher order quantities like mutual information~\cite{kraskov}, and divergence between two distributions~\cite{qingwang}. Below we make use of a generalization of the approach that allows binless estimation of the more nuanced  transfer entropy~\cite{paluscmi}. 

The basic idea behind non-parametric binless entropy estimators is to average local contributions to the entropy in the neighborhood of each point, where the neighborhood size is chosen adaptively according to the point's $k$ nearest neighbors.  The neighborhood shrinks as we add more data, improving the estimate. The fundamental strength of this approach comes from the fact that it is easier to locally estimate entropy than to locally estimate a probability density; mathematically, the equivalent local density estimators are not consistent while the entropy estimators are consistent~\cite{consistency}. For fixed $k$, various entropy estimators have been shown to be asymptotically unbiased and consistent under only mild assumptions~\cite{qingwang,kraskov,consistency,cmiconsistency}. 

Generally, suppose we have samples $i=1,\ldots,N$ of points $(\vec x^{(i)}, \vec y^{(i)})$ drawn from some unknown joint distribution. For each point, $i$, we construct the random variable $\epsilon_k(i)$, representing the distance to the $k$-th nearest neighbor in the joint $x$-$y$ space according to some metric. We will use the maximum norm in all dimensions following previous work~\cite{kraskov,paluscmi}.
For instance, the distance between points $i$ and $j$ in the joint space would be 
$$ \| \vec w^{(i)} -  \vec w^{(j)} \|_\infty = \max_l | w^{(i)}_l - w^{(j)}_l |,$$
where $\vec w^{(i)} = ( x_1^{(i)},\ldots,x_{d_x}^{(i)},y_1^{(i)},\ldots,y_{d_y}^{(i)})$ and $d_x, d_y$ are the dimensions of the $x$ and $y$ spaces, respectively.  
If we project only onto the $x$ (or $y$) subspace, the number of points strictly within a distance $\epsilon_k(i)$ is defined as $n_x(i)$ (or $n_y(i)$). 
We can now proceed to write down the Kraskov mutual information estimator~\cite{kraskov}.
\begin{eqnarray}\label{eq:mi}
\lefteqn{\hat H (X:Y) = \psi(k) +}  \\ && \frac1N \sum_{i=1}^N \left(  \psi(n_x(i)+1)+\psi(n_y(i)+1) - \psi(N) \right) \nonumber 
\end{eqnarray}
Here, $\psi$ is the digamma function.  Note that this simple expression depends only on distances between samples, and does not depend on the dimension of the space.

The Kraskov estimator has been extended to conditional mutual information(CMI)~\cite{paluscmi}. Now we add a third covarying vector, $\vec z$, and define $\epsilon_k(i)$ as the distance to the $k$-th nearest neighbor in the full joint $x$-$y$-$z$ space, while $n_{yz}(i)$, for instance, represents the number of points strictly within a distance $\epsilon_k(i)$ projecting onto the $y$-$z$ subspace. 
\begin{eqnarray}\label{eq:cmi}
\lefteqn{\hat H (X:Y|Z) = \psi(k) +} \\ && \frac1N \sum_{i=1}^N \left( \psi(n_{xz}(i)+1)+\psi(n_{yz}(i)+1)  - \psi(n_z(i)+1)  \right) \nonumber 
\end{eqnarray}
These two estimators can interpreted as estimators of time-delayed mutual information and transfer entropy, respectively, through appropriate choice of $X,Y,Z$. 

Note that because transfer entropy (and mutual information) estimators are an average over all samples, we can easily determine the contribution of one sample to the estimated entropy. 
\begin{eqnarray}\label{eq:lcmi}
\lefteqn{\hat H^{(i)} (X:Y|Z)  =\psi(k) +} \\ && \psi(n_{xz}(i)+1)+\psi(n_{yz}(i)+1) - \psi(n_z(i)+1)  \nonumber
\end{eqnarray}
We refer to this quantity as {\em local transfer entropy} and note its similarity to a previously introduced measure~\cite{lizier}. In principle, not only can we identify a pair of users, $X,Y$, so that $Y$ has high content transfer towards $X$, we can also order their tweet exchanges to see which ones contribute most to that assessment. An example is shown in Table \ref{tab:lcmi} in Sec.~\ref{sec:full}.

\subsection{Empirical study of CMI estimators}\label{sec:convergence}

Although entropy estimators have many nice theoretical properties in the asymptotic limit, for finite sample sizes we must ultimately rely on empirical results. Many papers have reported impressive empirical results from these entropy estimators already~\cite{victorbinless,qingwang,causalityreview,kraskov}, so we will explore only one unusual feature of our problem, with surprising results. 

\begin{figure}[htbp] 
   \centering
   \subfigure[]{
    \includegraphics[width=0.75\columnwidth]{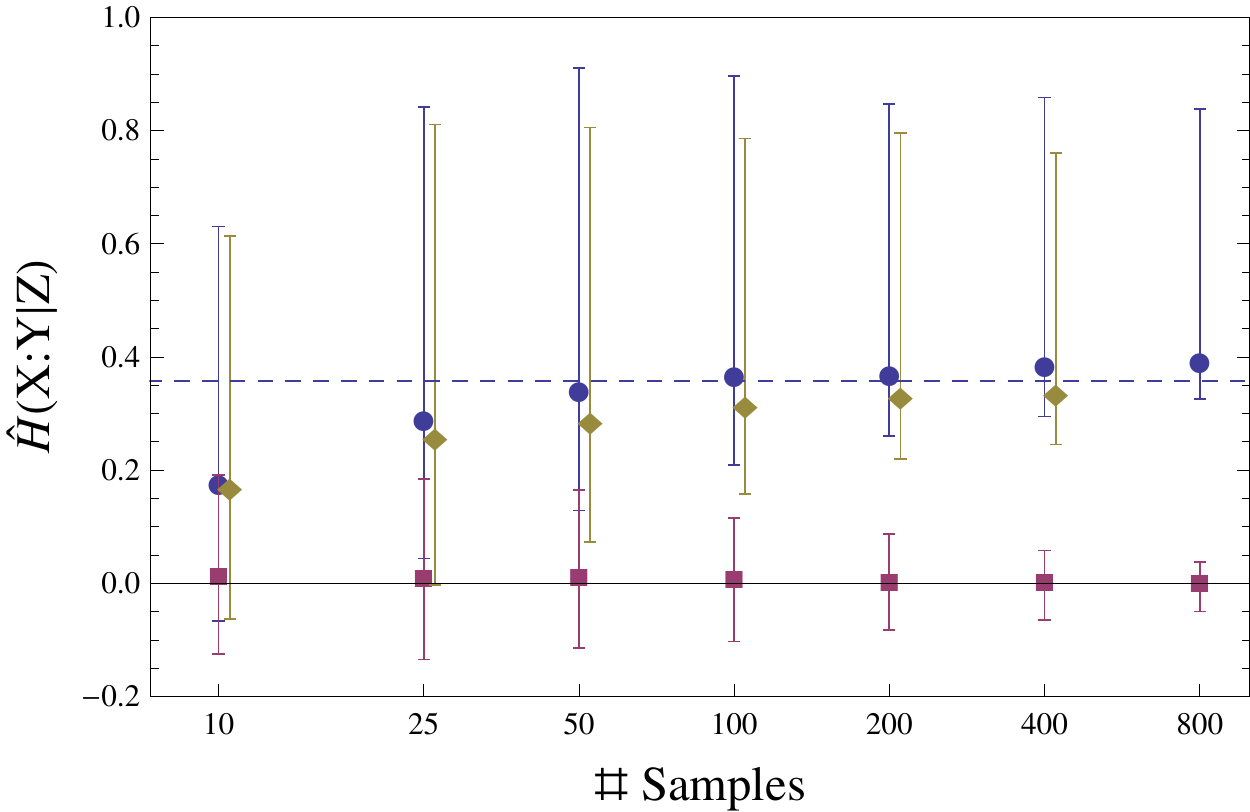} \label{converge}
    }
    \subfigure[]{
   \includegraphics[width=0.75\columnwidth]{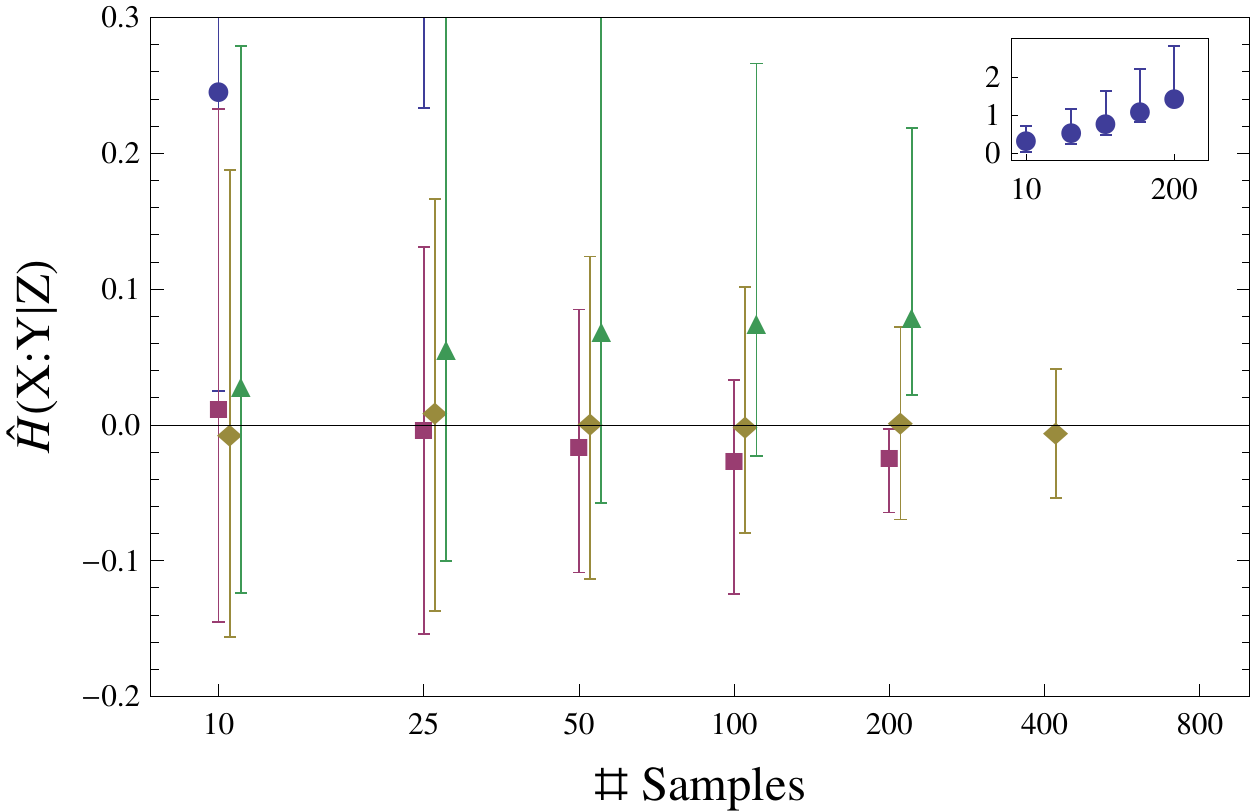} \label{converge2}
    }
   \caption{(a) Demonstrating convergence for the estimator in Eq.~\ref{eq:cmi}(circles) when the true CMI (dotted line) is known. This estimator succeeds even in the presence of many extraneous dimensions with small noise (diamonds). When labels are permuted, the estimator converges to zero (squares) (b) Comparing transfer entropy for pairs of interacting Twitter users (circles, continued in inset, and triangles) to null hypotheses (diamonds and squares). All points include $95\%$ confidence intervals, see Sec.~\ref{sec:convergence} for details. }
   \label{fig:test}
\end{figure}

Note that the estimators in Eq.~\ref{eq:mi} and \ref{eq:cmi}, do not explicitly rely on the dimension. In fact, they only rely on the vectors themselves through a distance function. Therefore, adding extra dimensions to a vector that are constant will have no effect on the distance function or the estimator. I.e., we would be transforming the vectors $\vec x^{(i)}$, 
$$ \vec x^{(i)'} = (x^{(i)}_1,\ldots,x^{(i)}_{d_x},c_1,\ldots,c_{d_c}),$$
where the $c_l$ are arbitrary constants (that is, they are the same for each point $i$). Clearly, the distances are unchanged, $|| \vec x^{(i)} - \vec x^{(j)}|| = || \vec x^{(i)'} - \vec x^{(j)'}|| $ (similarly in the joint $\vec x$-$\vec y$-$\vec z$ space), and this is all that is relevant for the estimators in Eq.~\ref{eq:mi} and \ref{eq:cmi}. The question we explore in Fig.~\ref{fig:test} is the effect of adding extra dimensions which are only nearly constant. The intuition for exploring this scenario is that vectors that represent content should be high dimensional, but we expect most individual users to participate in only a small subset of the full content space (we make this intuition concrete in Fig.~\ref{fig:ntopics} in Sec.~\ref{sec:topic}). Can we expect entropy estimators to work in this case?

We start by considering a situation in which we can calculate conditional mutual information(CMI) analytically. As an example, we consider a Gaussian where $\vec x, \vec y, \vec z \in \mathbb R^1$, and we take $X$ and $Y$ to be strongly correlated, while $Y,Z$ and $X,Z$ are weakly correlated. For illustration purposes, we consider the following covariance matrix.
$$\begin{pmatrix} x \\ y \\ z \end{pmatrix} \sim \mathcal N \left(\begin{pmatrix} 0 \\ 0 \\ 0 \end{pmatrix},\begin{pmatrix} 4 &3 &1 \\  3 & 4 & 1 \\ 1 & 1 & 2\end{pmatrix} \right)$$
In this case, $H(X:Y|Z) = 0.357$, while $H(X:Y) = 0.413$.\footnote{Both mutual information and conditional mutual information must be positive. For our purposes, conditioning on another variable generally reduces the mutual information, but it is possible for conditional mutual information to be larger than the associated mutual information.} Because $Z$ is correlated with $Y$, conditioning on it reduces some of $Y$'s usefulness for predicting $X$. In Fig.~\ref{converge}, we attempt to estimate this entropy from random samples. We show the mean of the estimator over many trials(circles), along with the $95\%$ confidence intervals. For comparison, we apply the estimator to samples where the sample indices of the $y$ and $z$ components are randomly permuted and it converges quickly to zero(squares). 

What happens if we now take $\vec x,\vec y, \vec z \in \mathbb R^{150}$? We will let $x_1,y_1,z_1$ be drawn from the same low-dimensional distribution above, but all the other components of the vector will be uncorrelated Gaussian noise with standard deviation $0.05$. The MI and CMI should be unchanged. Now we are estimating CMI in a 450 dimensional space with fewer than 400 samples. Surprisingly, the estimator still works well, only slightly underestimating the true CMI. If we had increased the standard deviation of the noise, eventually the signal would have been lost and the estimator would converge to 0. 

In Fig.~\ref{converge2}, we look at the convergence of the estimator for examples from pairs of users on Twitter (details in the Sec.~\ref{sec:results}). First, we consider a very strong signal corresponding to the edge $kar \rightarrow spo$ discussed later (circles). The estimate increases relatively quickly so that it must be continued in the inset. We consider two null hypotheses as comparisons. First, we permute the order of tweets for $kar,spo$ and calculate content transfer(squares). Second, we construct two Twitter streams from random tweets in our dataset, and we estimate the content transfer (diamonds). Finally, we calculate content transfer for the user pair $no\rightarrow li$ from Sec.~\ref{sec:mention}, which represents more social behavior (triangles). Note that the estimator in this case is quite noisy, and we do not expect perfect discrimination of the signal with so little data. 

\subsection{Implementation details}\label{sec:implementation}
There are several details to be considered before implementing the estimators above. First of all, we are required to find the $k$-nearest neighbors to each point, but how should we choose $k$? Smaller $k$ reduces the bias, but larger $k$ reduces the variance~\cite{qingwang}. We find the results are not very sensitive to $k$. We use $k=3$ as suggested by Kraskov et al.~\cite{kraskov} and this choice is confirmed by numerical results shown in the inset of Fig.~\ref{fig:mentions}. To avoid situations where two points are exactly the same distance away, we also add low intensity ($10^{-10}$) noise to the data~\cite{kraskov}. 

The most intensive part of the calculation is the search for nearest neighbors. In high dimensions, as is the case in this paper, this cannot be sped up much beyond $O(N^2)$ for $N$ samples. However, we can fix some constant $N_c$, and only make estimates using samples of this fixed size. Besides bounding the computational complexity, if we average over multiple samples we can also reduce the variance. For $N$ samples of tweet exchanges, we take $2 \lceil N/N_c \rceil$ random subsets of size $N_c$. 
We set $N_c = 100$ (which will be the minimum sample size we keep in our data, discussed in Sec.~\ref{sec:results}). This also insures that any bias from finite sample size will affect all edges equally.  Because we attempt to evaluate transfer entropy between all of the millions of user pairs (only some of which have the minimum number of samples), we had to split our calculation over many processors.\footnote{All results were obtained on USC's HPCC~\cite{hpc}. Code is available: \url{http://www.isi.edu/~gregv/te.py} }

\section{Results}\label{sec:results}
We will apply the entropy estimators to real world data from Twitter. After describing the dataset, we will discuss options for representing tweet text as vectors. In Sec.~\ref{sec:full}, we will examine the directional links with the highest content transfer on the entire dataset. Although we lack a ground truth to validate our results, in Sec.~\ref{sec:mention} we consider activity of a subset of users for whom mentions can be used to test the significance of content transfer. 


\subsection{Dataset description}
We make use of data originally collected and described in \cite{macskassy2012}. All tweets are collected for a set of 2400 users over a one month period from 9/20/2010--10/20/2010. The set of users was picked by starting with a small, random initial set and constructing a snowball sample using mentions and re-tweets. The dataset was constrained to users who self-reported in their profile a location in the Middle East. The dataset also contained sampled tweets from tens of thousands of other users who mentioned users in the original set. We used those tweets to help train the topic model (giving a total of over half a million tweets after preprocessing described in the next section), but we did not consider those users when calculating content transfer for pairs of users. 
After eliminating users with less than 100 tweets, we considered all possible directed edges among the remaining 770 users. Note that not all ordered pairs of users had at least 100 tweet triples as defined in Sec.~\ref{sec:preliminaries} in which case content transfer was not calculated.

\subsection{Topic vector representation}\label{sec:topic}

A crucial ingredient in our attempt to apply information-theoretic measures to social media is a way to represent content as vectors. Luckily, a great deal of work has been done on mathematical representations of content, a few examples are discussed in Sec.~\ref{sec:related}. The richness of our results are ultimately limited by the quality of content representation. On the other hand, higher dimensional representations make entropy estimation more difficult. 

Compounding this difficulty, social media presents some unique challenges. For instance, on Twitter, messages are very short (140 characters), providing little context to determine what a tweet is about. The use of ``netspeak'', emoticons, and abbreviations challenge traditional models of communication. Many languages are represented, sometimes mixed within a single tweet. Spelling mistakes and URLs multiply the number of unique tokens. On the other hand, the sheer volume of data provides an advantage that outweighs these difficulties.

Ultimately, we chose to use topic models to represent tweets because of convenient off-the-shelf implementations\cite{gensim}, and a growing body of work exploring their applicability to Twitter\cite{smutopic,twittertopic}. 
Our purpose in using a topic model differs from standard aims. In particular, our ultimate goal is not to find distinct topics with clear interpretations, but to find the minimal representation that preserves relevant detail.

For our experiments, we trained an LDA topic model implemented in {\em gensim}\cite{gensim}.
For pre-processing the text, we followed most of the prescriptions in \cite{twittertopic}.  (1) We replace all URLs with the word ``[url]''. (2) We replace all words starting with ``@'' with the word ``[mention]'' (3) We remove all non-Latin alphabet characters and convert to lower-case. (4) We removed a standard list of English stop-words. Because we will use mentions later in our validation, step (2) is particularly important to insure that our topic model has not learned name associations. We also removed all tweets that begin ``RT @'' (re-tweets), since this type of information diffusion has been well-studied, as discussed in Sec.~\ref{sec:related}.

\begin{figure}[htbp] 
   \centering
   \includegraphics[width=0.8\columnwidth]{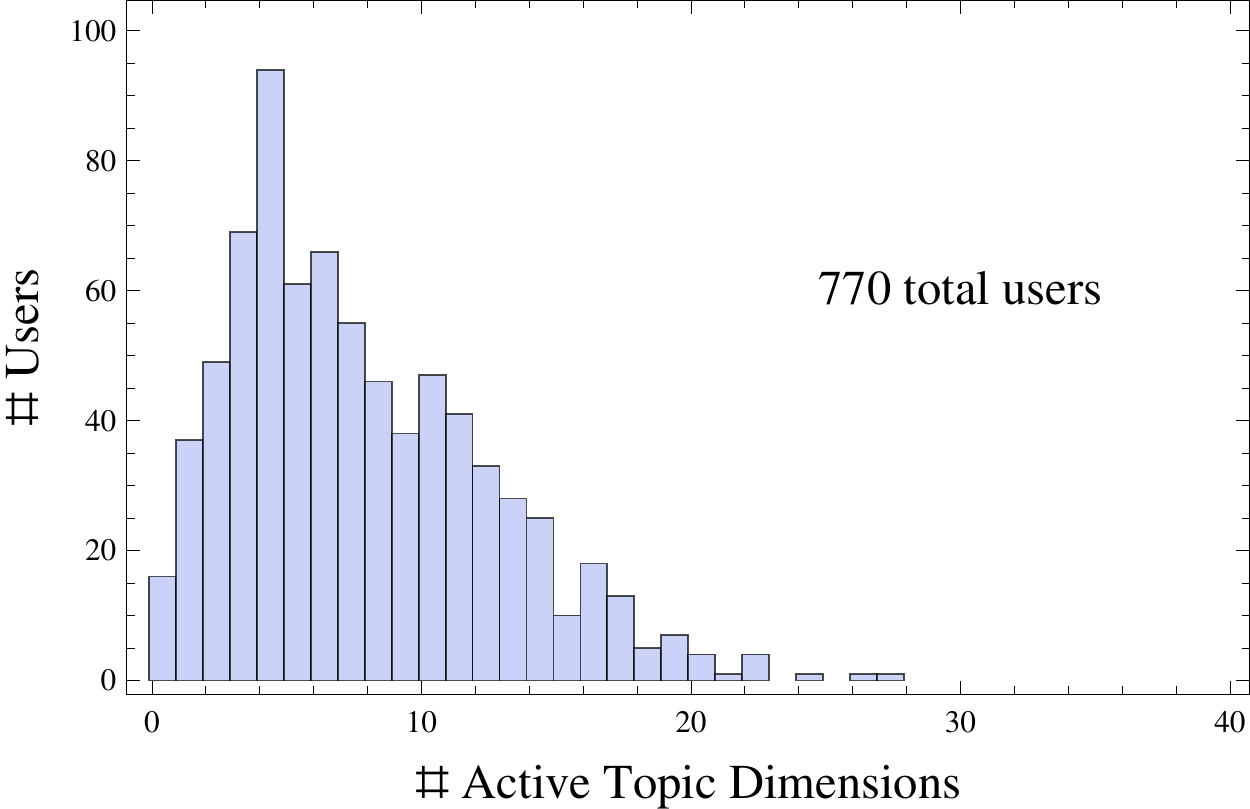} 
   \caption{The effective dimension of topic vectors for most users is low. We consider all users with at least $100$ tweets. For each component of the $150$ dimensional topic vector, we calculate the standard deviation over all tweets from one user. If the standard deviation is greater than $0.05$, we say that topic dimension is ``active'' for that user.}
   \label{fig:ntopics}
\end{figure}
Next, we constructed a bag-of-words vector representation with the remaining words in the dataset that appeared more than once. Each component of the vector represents the frequency with which one of these words occurred in a given tweet. These vectors were transformed using the TF-IDF score.\footnote{Term frequency-inverse document frequency: we use the standard definition. If a term occurs $f$ times it is transformed to $f \log_2 D/d$, where $D$ represents the number of documents and $d$ represents the number of documents containing the term.} Finally, we used the TF-IDF vectors to learn an LDA topic model. We tried topic models with 10, 50, 100, 125, 150, 175, and 200 topics. At first, we assumed that the lower dimensional topic models, while being worse representations of the text, would be more amenable to entropy estimators. However, larger topic models actually fared much better, despite the high dimensionality of the vectors. One reason for this surprising result is that the effective dimensionality of most Twitter users is far smaller than the dimensionality of the topic vector. To verify this, we considered all the users in our dataset with at least 100 tweets. For each component of the topic vector, we calculated the standard deviation over all tweets for one user. We define the number of ``active topics'' as those for which the standard deviation was over 0.05.  Although this notion does not conform to a standard intuition about what should be considered an ``active topic'', it does describe what is relevant for entropy estimation, as discussed in Sec.~\ref{sec:convergence}. 
The result is shown in  Fig.~\ref{fig:ntopics} for a topic model with 150 topics. The dynamics of most users are constrained to a handful of dimensions. The fact that the active topics may differ for different users is irrelevant for the purpose of entropy estimation.

\subsection{Full graph}\label{sec:full}

\begin{figure}[htbp] 
   \centering
   \includegraphics[width=0.95\columnwidth]{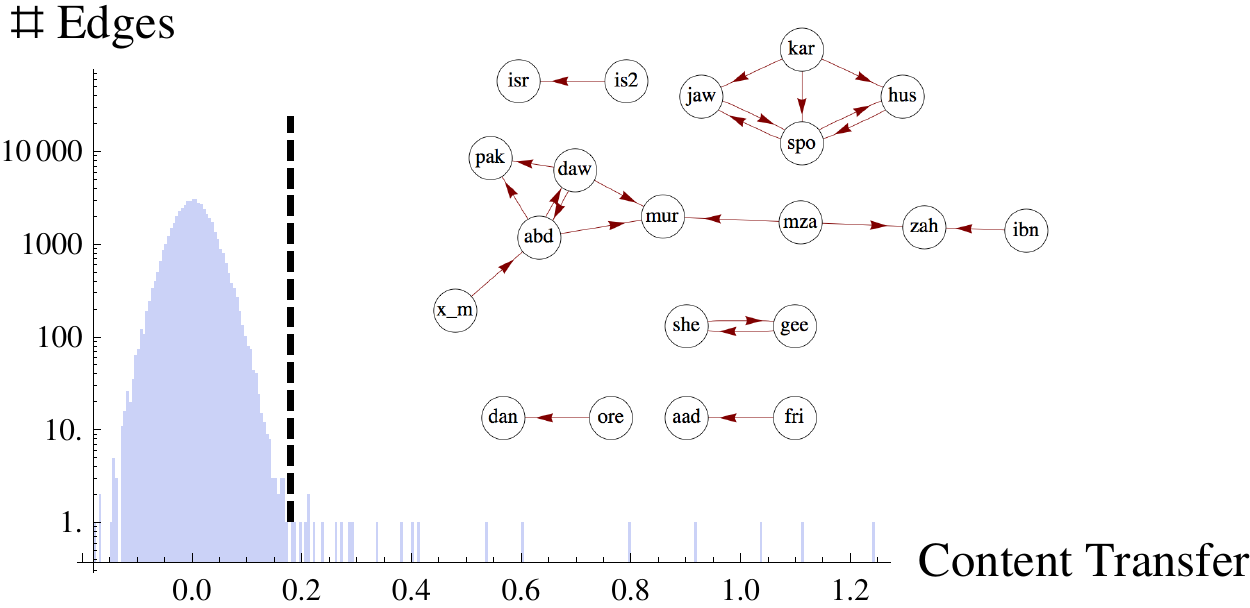} 
   \caption{Histogram of content transfer for all ordered pairs of users with sufficient samples. Edges with content transfer to the right of the dotted line are shown in the inset graph. Many of these edges are not present in the follower graph or mention graph. }
   \label{fig:tehistogram}
\end{figure}

We begin by calculating content transfer for all ordered pairs of users with sufficient samples. Unless otherwise specified, we use $n_{topics} = 125$ in the following examples, although the high content transfer edges were insensitive to this choice. Looking at the histogram of content transfer for all edges in Fig.~\ref{fig:tehistogram}, we note that there are a few obvious outliers. We also show the network consisting of only these high content transfer edges, with account names abbreviated. Inspection of the tweets reveal that these links are all strongly predictive, we proceed to give several examples.

We start with the edge $mza \rightarrow zah$ with sample tweets shown in Table \ref{tab:lcmi} and point out some things notably lacking. For these two users there are no friend links, no mentions, no retweets, no matching URLs (though the shortened URLs point to the same stories), and no matching hash tags. Nevertheless, even though the text is altered, content transfer recognizes that $zah$'s tweets are identical in content to $mza$'s tweets. We use the local transfer entropy from Eq.~\ref{eq:lcmi} to order the 288 tweet exchanges. 
We also read through all the tweet exchanges and hand labeled 228 instances in which the two users' tweets clearly referred to the same story. The probability that the local transfer entropy was higher for a tweet exchange which was a duplicate than for a non-duplicate one was 0.68. 
We also note the asymmetry of this edge: the content transfer from $mza \rightarrow zah$ is $0.24$ while  in the other direction it is only $0.01$. 
This asymmetry is often taken to suggest a causal connection\cite{paluscmi}. 
Nevertheless, there remains the possibility of an external, mutual cause. The impossibility of ruling out such alternatives is one reason we emphasize the interpretation of content transfer as a measure of predictability. Of course, this is a well-known caveat regarding Granger causality. 
A simple explanation in this case is that both accounts simply read and post from the same news site. 
However, in that case we would expect the order of tweets to sometimes be reversed, causing the transfer entropy to be more symmetric. In fact, the order of tweets is always preserved. A more nuanced alternative is that one of the users is temporally ``closer'' to the news source. E.g., a service like ``twitterfeed.com'' can automatically post news stories to your Twitter account the instant they are published.

As opposed to the previous example, the $kar$ cluster contains some bi-directed edges. In this case, the users are all following each other, however, once again, no retweets or mentions are used. The tweets revolve around sports and some samples are shown in Table \ref{tab:sports}. The tweets are clearly all copies of each other. Confirming the previous intuition about temporal ordering, the bi-directed edges are duplicates that occur in arbitrary order, while the directed edge away from $kar$ is reflected by the fact that all posts appear first on that account. The account $hus$ appears to be a personal account that occasionally included unrelated tweets. The profile of that account describes the author as a ``sports analyst.'' 

The remaining clusters in Fig.~\ref{fig:tehistogram} have similar easy qualitative interpretations. The $isr$ cluster users post identical Israeli news stories. The remaining edges of the largest cluster also revolve around news, mostly of stories in the Middle East.  The profile of $gee$ lists itself as the Twitter stream of a tech news site, while the profile of $she$ lists itself as the founder of the same website. Tweets are copied in arbitrary order, leading to symmetric content transfer. The edge $aad \leftrightarrow fri$ has a similar interpretation, with $aad$'s profile declaring himself a radio presenter for the internet radio station represented by account $fri$. Again, no mention or follower edges are declared. 

\begin{figure}[htbp] 
   \centering
   \includegraphics[width=0.8\columnwidth]{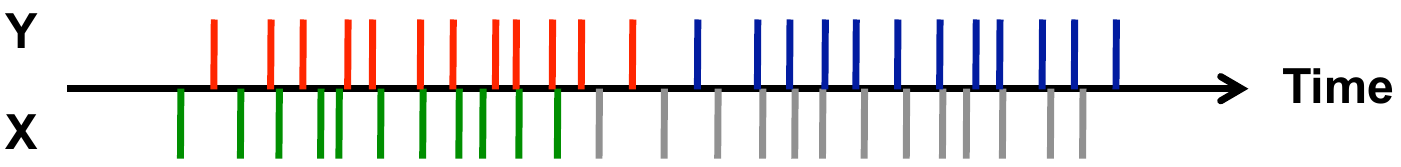} 
   \caption{Illustration of a scenario when time delayed mutual information is high while transfer entropy is low. Colors represent repetition of a particular tweet.}
   \label{fig:cmi_v_mi}
\end{figure}
We also calculated the time-delayed mutual information for all pairs of users. In general, this quantity was correlated with the content transfer (see Fig.~\ref{fig:mentions}). However, there were several examples where mutual information was high while content transfer was low. The intuition behind this phenomena is expressed in Fig.~\ref{fig:cmi_v_mi}. To use a concrete example from the dataset, we have user Y repeatedly tweeting the same message 
\texttt{\#shoutout for Floods in \#Pakistan \#pkfloods [url]}, 
which we can imagine as a red line in the picture. At some point, user Y switches to repeatedly tweeting a new message 
\texttt{\#TEAMFOLLOW 100 FREE MORE TWITTER FOLLOWERS! [url]}\footnote{This tweet refers to increasingly popular services to inflate the number of followers for an account. Basically, you agree to either pay, follow people in the service, or tweet advertisements about the service in exchange for followers. This service formed a prominent signal in \cite{versteeg2012www}.}
 (blue lines). Coincidentally, at nearly the same time user X switches from repeatedly tweeting 
 \texttt{In My View: HOW TO EXPLODE RESOURCES TO EARN FOREIGN EXCHANGE [url]}
  (green lines) to 
  \texttt{In My View: PAKISTAN ON THE SHOULDERS OF N.R.O BENEFICIARIES [url]}\footnote{We also estimated the mutual information $\hat H(X^{ \mbox{\tiny F}}:X^{ \mbox{\tiny P}})$, and this user was an example that had a high score for this measure. Basically, the user tweets many different story titles, but repeats each one dozens of times. }
   (grey lines). From looking at the figure, you can see that a red line for user Y is quite predictive of a subsequent green line for user X, and this is reflected in the high mutual information. On the other hand, if you condition on user X's past, you can easily predict that a green line is most likely followed by another green line, and seeing a red line from user Y does not improve that prediction. Therefore, transfer entropy is low in this scenario.

\begin{table}[htbp] \tiny
\centering
\begin{tabular}{|l|l|p{6cm}|} \hline
LTE & User & Tweet\\
2.65 &
zah& KARACHI, Pakistan, Oct. 12 (UPI) -- Intelligence agencies in Pakistan are warning of terrorist atta... http://bit.ly/bscYoX \#news \#Pakistan\\
&mza& Is Mobile Video Chat Ready for Business Use?: Matthew Latkiewicz works at Zendesk.com, creators of web-based custo... http://bit.ly/cAx3Ob \\
&zah& Matthew Latkiewicz works at Zendesk.com, creators of web-based customer support software. He writes for... http://bit.ly/bkuWCV \#technology \\ \hline
2.53 &
zah& Man-made causes cited for Pakistan floods: ISLAMABAD, Pakistan, Oct. 14 (UPI) -- Deforestation ... http://bit.ly/92afA0 \#pkfloods \#Pakistan \\
&mza& Google Shares Jump 7\% on Impressive Earnings: Google has posted its latest earnings report, and early indications ... http://bit.ly/9oi4zr \\
&zah& Google has posted its latest earnings report, and early indications suggest that investors are more tha... http://bit.ly/cyT35p \#technology \\ \hline
-0.33&
zah& ISLAMABAD: Former adviser Sardar Latif Khosa resigned as adviser to the prime minister on Tuesday giv... http://bit.ly/bBKJ32 \#news \#latest \\
&mza& Explore Interesting, Personal Photos on Yahoo! Search: Following up on our announcement of new Yahoo! Search enhan... http://bit.ly/diR9ls \\
&zah& KABUL: A cargo plane crashed into mountains east of Afghanistan's capital Kabul on Tuesday, with init... http://bit.ly/9ExkjR \#news \#latest \\ \hline
\end{tabular}
\caption{Tweet exchanges between two users with high content transfer from $mza \rightarrow zah$. Examples were picked which have high and low local transfer entropy.}
\label{tab:lcmi}
\end{table}

\begin{table}[htbp] \tiny
\centering
\begin{tabular}{|l|p{6.5cm}|} \hline
kar& [url] -- South Korean GP gets F1 go-ahead \\
hus& [url] -- South Korean GP gets F1 go-ahead \\
jaw& [url] -- South Korean GP gets F1 go-ahead \\
spo& [url] -- South Korean GP gets F1 go-ahead \\
\hline
kar& The Sports Encounter -- It's all about Sports -- ICC cleared Oval ODI [url] via @AddThis\\
jaw&The Sports Encounter -- It's all about Sports -- ICC cleared Oval ODI [url] via @addthis\\
spo& The Sports Encounter -- It's all about Sports -- ICC cleared Oval ODI [url] via @AddThis\\
hus& The Sports Encounter -- It's all about Sports -- ICC cleared Oval ODI [url] via @AddThis\\
\hline
\end{tabular}
\caption{The sports cluster in Fig.~\ref{fig:tehistogram}. User $kar$ always tweets first, while the others repeat sporadically and in varied order. }
\label{tab:sports}
\end{table}



\subsection{High mention users}\label{sec:mention}

We saw that the user pairs with the highest transfer entropy, while clearly good predictors of content, did not have the typical markers associated with ``influence'': no mentions, no retweets, and no following. In this section we consider a subset of Twitter users with two goals. (1) Can content transfer capture more subtle social behavior? (2) Can we use some of the metadata about mentions and followers to validate the effectiveness of content transfer? 

\begin{figure}[htbp] 
   \centering
   \includegraphics[width=0.9\columnwidth]{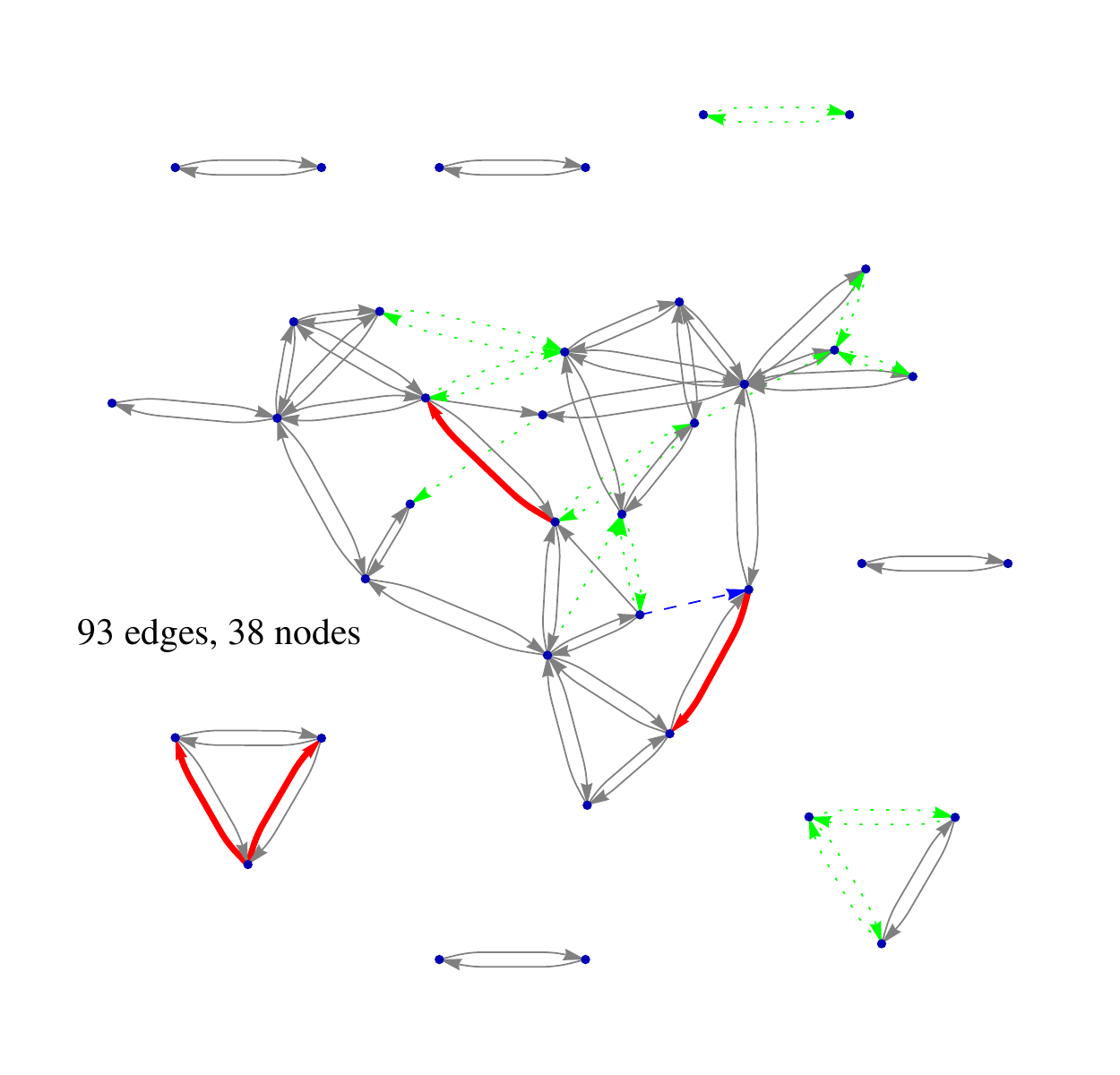} 
   \caption{The mention graph for users described in Sec~\ref{sec:mention}. The five edges with the highest transfer entropy are highlighted. True positives are thick red arrows and false positives are blue dashed arrows. The green, dotted arrows are mention edges for which content transfer was not calculated due to insufficient data.}
   \label{fig:mentiongraph}
\end{figure}

A first idea would be to consider the set of users who use ``@mentions'' with the intuition that if a user mentions somebody, they are responding to them somehow. Unfortunately, this intuition would often fail. A major use of mentions is to attempt to get the attention of celebrity Twitter users. E.g. one user mentioned ``@justinbieber'' over seven hundred times during the month of our dataset. Instead we want to find people who use mentions for conversation, which was the primary focus of Macskassy~\cite{macskassy2012}. To do this, for each pair of users, we call the number of mutual mentions the minimum of the number of times $X$ mentions $Y$ and the number of times $Y$ mentions $X$. We ranked the top 50 pairs with the highest mutual mentions. For this set of users (38 users total), we can reasonably assume that mentions are a weak proxy for online influence. The full mention graph for this set of users is shown in Fig.~\ref{fig:mentiongraph}. 

\begin{figure}[htbp]
\centering
\subfigure[]{
    \includegraphics[width=0.8\columnwidth]{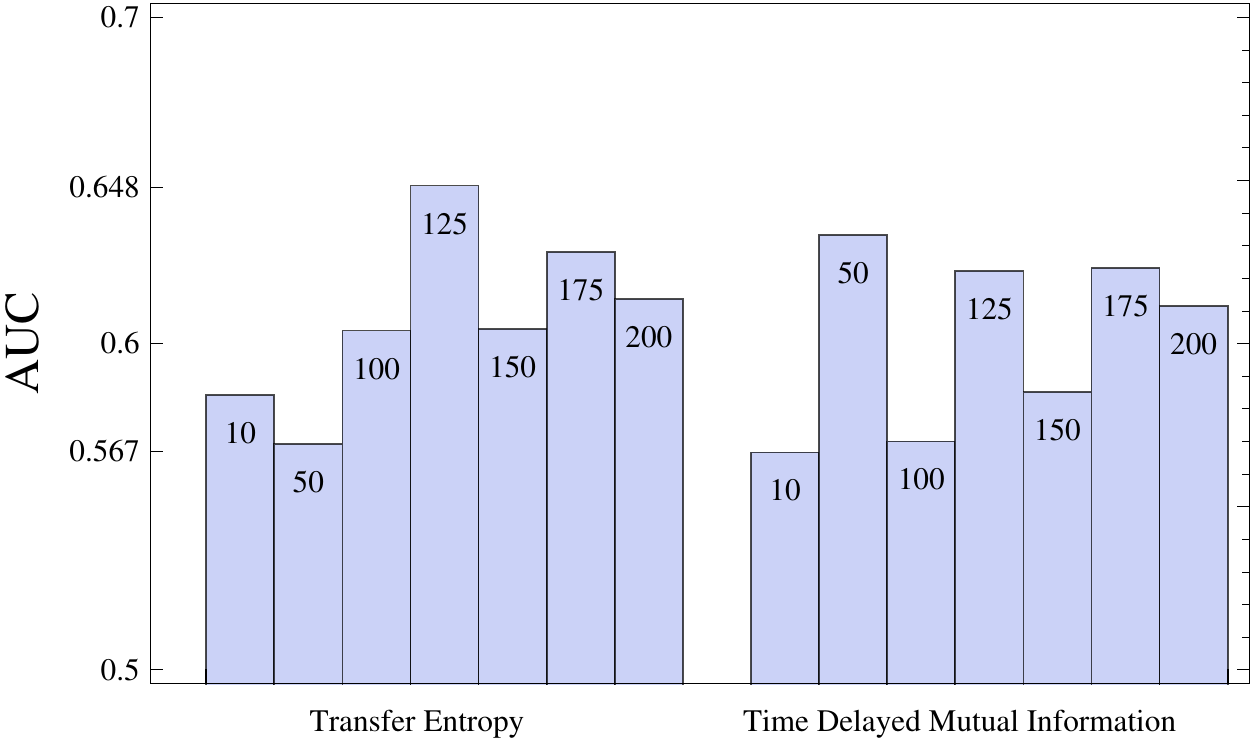} \label{cmi_mi}
    } 
    \subfigure[]{
    \includegraphics[width=0.8\columnwidth]{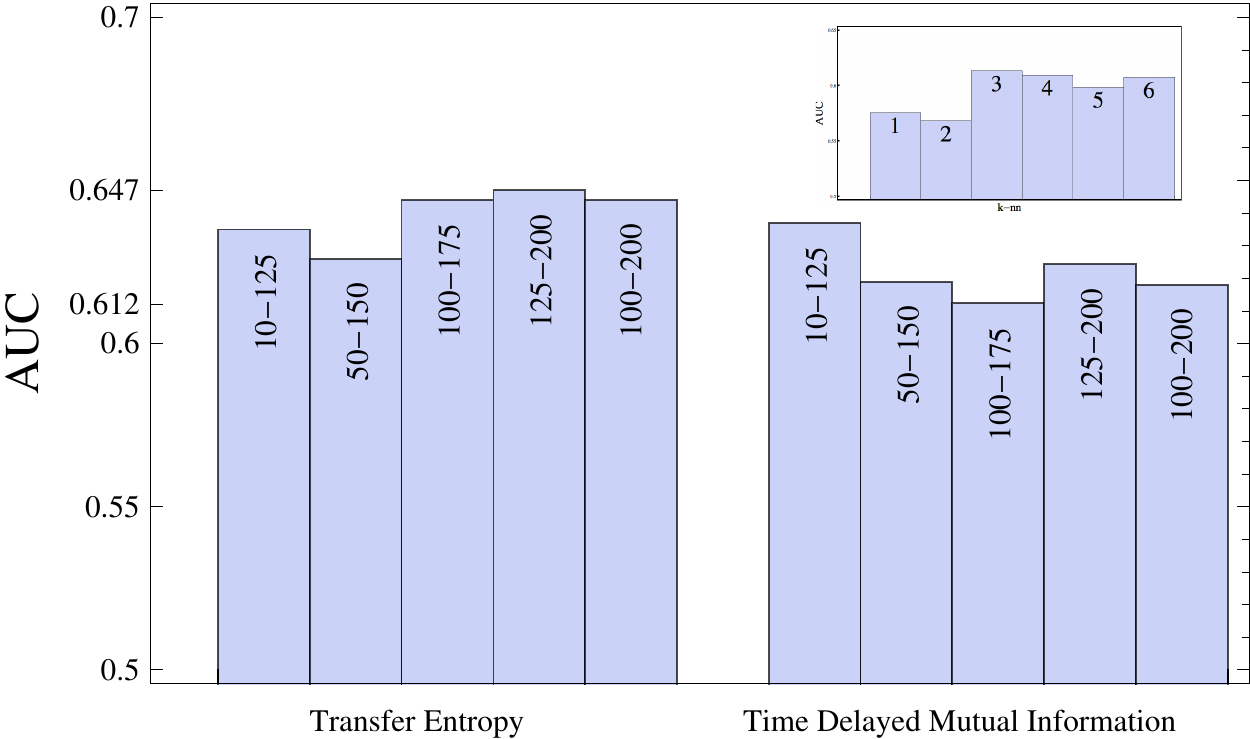} \label{cmi_mi_avg}
    }
     \caption{AUCs for reconstructing the mention graph in Sec.~\ref{sec:mention} for various parameters. (a) Transfer entropy and MI for various topic models, $k=3$. (b) AUC using the average ranking for several topic models. Inset shows the effect of changing $k$, for a topic model with 100 topics.  }
   \label{fig:mentions}
\end{figure}

For this restricted set of users, we see how well content transfer corresponds to the underlying mention graph. For an evaluation metric we use area under the receiver operating characteristic curve (AUC)~\cite{auc}. We rank all edges according to which have the highest content transfer. In principle, we would like the edges in the mention graph to have the highest content transfer. The AUC can be interpreted as the probability that a mention edge ($X$ mentions $Y$ at least once) has a higher content transfer ($IT_{Y \rightarrow X}$) than an edge without a mention. As a null hypothesis, we can imagine that content transfer scores are random. In that case the mean AUC will be $0.5$. Since we have 74 mention edges and 785 non-mention edges, the standard error of the AUC under the null hypothesis is about $3.5\%$~\cite{auc}.  The best result for AUC (using $n_{topics}=125$) in Fig.~\ref{fig:mentions} is 0.648, which is over four standard deviations away from the null hypothesis.  As an alternate perspective, we point out that the precision and recall for the top 100 edges were $20\%,28\%$, respectively, which are both about twice the baseline. 

Fig.~\ref{cmi_mi} shows the AUC for various topic models using either transfer entropy or time-delayed mutual information. Because the results are noisy and, in principle, our method does not rely on the details of any particular topic model, in Fig.~\ref{cmi_mi_avg}, we used the average rank given by multiple topic models to predict mention edges. On average, transfer entropy slightly outperforms time-delayed mutual information. The inset of Fig.~\ref{cmi_mi_avg} shows the effect of trying various $k$ (number of nearest neighbors in our entropy estimator), with $n_{topics} = 100$. We see that $k=3$ is a good choice, but larger values of $k$ give similar results.

In Fig.~\ref{fig:mentiongraph}, we consider the full mention graph of the set users with high mutual mentions. We also highlight the top 5 edges according to content transfer for $n_{topics} = 125$ since it gave the best AUC. Four edges are true positives, with example tweets shown in Table \ref{tab:mentions}. Some comments are in order about how tweet processing affected these examples. First of all, we eliminated messages that started with ``RT @'', but not ``partial retweets'', where a message was prepended to the re-tweet. Second, no language detection was done, though eliminating non-Latin characters eliminated most of the foreign text which was in Arabic and Hebrew. However, a tweet containing any Latin characters (including mentions) was still represented. 
Most of the topics represented English words, but a few topics contained mostly transliterated Hindu and Urdu, or Spanish. Identifying that if one user tweets in a language another user will respond in kind is a strongly predictive signal. 

The discussion in Table \ref{tab:mentions} between $sh$ and $ta$ perhaps represents the ideal for detecting social influence. In that case we see that if one user broaches a political topic, the other will follow suit. Note that this pair of users was detected through content dynamics alone, without reference to mentions or even knowledge of follow edges. In fact, the distance is not even calculated between tweet vectors for $sh$ and $ta$, the content transfer only looks for a predictable {\em co-variation} in the content of their tweets.

\begin{table}[htbp] \tiny
\centering
\begin{tabular}{|l|p{6.5cm}|} \hline
User & Tweet\\
\hline
sh & @ta tsalk to police officers. 6 prominent policemen of Op Cleanup have been killed in last 2 yrs. Still tolerating MQM \\
ta & @sh I meant the "participation" of the hijacked public was a function of fear perp by Talibs. Same thing here. ppl don't want 2 die \\
sh & @ta what does it serve them?More pathetic f*tards snatching their mobiles and wallets? Small-crime is engrained in MQM structure \\
ta & @sh re: "no soul n honor"... well I think MQM zia's creation to puncture the Sindh Nationalist cause. ISI \_will\_ slap its b* \\
\hline
\hline
en & @fz oye oye Ajj MJ ke barri hai .. tm kal sa lyrics tweet ker rahe ho .. khariat hai na larki ? \\
fz & @EN Jo b dimagh mai ata hy krdaiti hun m not listening to him atm though \\
en & @fz mujhe na urdu na english songs ka lyrics kabhe yadd howa hain ...bhalla bhae tera :P \\
fz & @EN Lol i love memorizing songs ;) \\ \hline
en & NEW VERSION OF TWITTER IS HERE ... \\
fz & u got it? :O RT @EN NEW VERSION OF TWITTER IS HERE .. \\
en & @fz YEAH I THINK SO ... YOU GOT IT ??? SPLIT SCREEN VERSION ? \\
fz & @EN no :( m still waiting for it \\
\hline
\hline
re & queremos unaa fotooooo deee @celeb1 y @celeb2 \\
li & QUIERO UNA FOTO DE @celeb1 \& @celeb2 \\
no & @celeb2 nico .. please que la segunda imagen sera de vos con @celeb1 \\
re & duele tanto decir ALGO ? \\
li & @celeb2 nico porfi saca una foto con emi :( \\ \hline 
re & @No [Hebrew characters] \\ 
no & @Li @Re [Hebrew characters] \\ \hline
no & @re twiitcam baby, yes o no?! \\
re & @No yesssss, and my brother will be theirr !! hahah , your sweet \\
no & @Re jaja! very good sister! :) \\ \hline
\end{tabular}
\caption{Representative examples of tweet exchanges between the four pairs of users identified as being among the top 5 highest transfer entropies for all user pairs defined in Sec.~\ref{sec:mention}. Edges were $no\rightarrow re$, $no \rightarrow li$, $en \rightarrow fz$, $ta \rightarrow sh$. User's $re,no,li$ tweet primarily in Spanish, but all three occasionally address each other and respond in Hebrew or English. }
\label{tab:mentions}
\end{table}

\section{Related work}\label{sec:related}


Much research has focused on characterizing and identifying influential users that can facilitate information diffusion along social links. Researchers have suggested different characterizations of influentials based on various network centrality measures~\cite{PageRank,Jeh03,Ghosh10snakdd}. For Twitter data, various influence measures include number of followers, mentions, retweets~\cite{Cha10measuringuser}, Pagerank of follower network~\cite{Kwak2010}, size of the information cascades~\cite{Bakshy11}. More recent work has attempted to utilize temporal information through  the influence--passivity score~\cite{Romero2010}, and transfer entropy~\cite{versteeg2012www}. None of those measures, however, take content into account. More recently, several authors have suggested topic-sensitive influence measures such as TwitterRank~\cite{Weng2010WSDM}, which takes into account topical similarity among the users. Topic-specific re-tweeting behavior was examined in~\cite{Macskassy2011ICWSM}.

More generally, there is an increasing trend to use communication content for inferring the nature of relationships between users~\cite{Bramsen2011HLT,Diehl2007AAAI,Gilbert2009CHI}. 
An interesting  line of  research grounded in psycholinguistic theory of communication have studied  the convergence of communicative behavior among Twitter users~\cite{Danescu-Niculescu-Mizil2011WWW}. In particular, it has been suggested that conversational behavior can be indicative of relative social status of participants, and subtle language-based signals can be used to infer  power relationships among the users~\cite{Danescu-Niculescu-Mizil2012WWW}. Similar to our work, those approaches too work by projecting unstructured user-generated text  onto a multivariate time series, in their case using LIWC categories~\cite{Pennebaker} rather than LDA-induced topics. However,  the influence measures suggested in~\cite{Danescu-Niculescu-Mizil2011WWW,Danescu-Niculescu-Mizil2012WWW} are defined in a rather {\em ad hoc} manner, as opposed to a more fundamental entropic measure used here. We believe that our approach based on transfer entropy provides a more principled measure of directed influence.

A crucial component of our approach  is based on the ability to estimate entropic quantities for very-high-dimensional random variables. Due to data sparsity, naive methods based on {\em binning}  are not feasible. The binless approach for entropy estimation introduced in~\cite{Kozachenko} has been used for quantifying information in neural spike trains~\cite{victor}. The binless approach has been extended for estimating higher order entropic quantities such as mutual information~\cite{kraskov}, divergences between two distributions~\cite{qingwang}, and transfer entropy~\cite{paluscmi}.  We also note that a linear version of the transfer entropy known as Granger causality~\cite{Granger1980} has been used recently for uncovering predictive causal  relationships in neuroscience~\cite{Kaminski2001}, genetics~\cite{Lozano2009}, climate modeling~\cite{LozanoKDD2009} and various other applications.

\section{Discussion}\label{sec:conclusion}

We have seen that using content transfer as a general, statistical measure of predictivity captures a wide variety of nontrivial behavior on Twitter. Information-theoretic techniques provide powerful, flexible tools for discovering patterns in data, but typically are impractical to implement. Surprisingly, non-parametric entropy estimation was quite effective on a dataset that would be considered small by recent research standards. This is despite the fine-grain application of these entropic measures to individual user pairs. Extraordinarily, Table~\ref{tab:lcmi} suggests the measure may even provide a meaningful signal at the level of individual tweets. 

The strongest, most predictive signals discovered in Sec.~\ref{sec:full} were all characterized by some type of news dissemination. 
Most interesting about these results were how many of the links appeared to be purposely hidden in the explicit follower graph.
If news dissemination is for the purpose of promoting your internet radio station, as in the $fri \leftrightarrow aad$ example, it may be advantageous for the accounts promoting your web site to appear as independent as possible. Indeed, Twitter terms of usage prohibit automatic re-tweets, so if you are copying content on multiple accounts, it would be a mistake to call attention to the practice by using re-tweets.
Ironically, for these purposes it may be advantageous to hide the truly influential edges, while at the same time it is advantageous to accrue as many followers as possible to appear influential, even if most of these followers are dummy accounts that are not influenced at all. 

We also found a statistically significant result in Sec.~\ref{sec:mention} for distinguishing ``social influence'', i.e., one user eliciting a response in another. The evaluation task we performed is akin to hearing hundreds of people talking at once and inferring who is talking to whom, just by the content of their statements and without reference to any explicitly declared relationships. While the data were not sufficient to distinguish an arbitrary social tie, on average edges identified with mentions had a higher content transfer and this effect was over four standard deviations from the null hypothesis.
One of the top examples corresponded to an intuitive notion of social influence, revolving around political discussion, but the strongest signals were for multi-lingual users. Responding in-kind to a certain language is a relatively easy signal to identify, at least within a topic model representation. 
We can see from Fig.~\ref{fig:test} that to distinguish independent signals from correlated ones with transfer entropy requires either a strong signal or more data. It would be interesting to see what types of social influence are detected with even an order of magnitude more data, which is still many orders of magnitude away from the amount of data regularly processed by companies like Twitter and Facebook. 

A subtle point about our measure of content transfer is that at no point are $Y$'s tweets directly compared to $X$'s tweets. Rather, the measure checks if $X$'s future content varies in a predictable way based on $Y$'s past content. While this distinction may be overly general for the purpose of discovering connections from what topics are being discussed, it may be relevant for more subtle social cues. For instance, whenever $Y$ makes an aggressive statement, if $X$ always responds submissively, this is a predictable, but not matching, response. To capture this type of scenario would require content representation that includes things like stance, attitude, or sentiment, as discussed in Sec.~\ref{sec:related}. 

Social media is in a state of constant growth and change. Subtle changes in the mechanisms that underly social media platforms can have dramatic effects on the user behavior~\cite{hodas2012visibility}. Results based on detailed modeling of hash tags, mentions, or re-tweets may not be relevant for the next generation of social media. 
On the other hand, a measure based on information-theoretic principles will remain relevant for any communication medium. 
On a more practical note, by providing a model-free way to discover unexpected relationships in data, information-theoretic analysis is an effective tool for data exploration. 

\subsection*{Acknowledgments}
We would like to thank Sofus Macskassy for providing data and valuable comments. We also thank Kristina Lerman and Jaebong Yoo for useful discussions. GV would like to thank LARC, SMU for their hospitality while finishing this work.
This research was supported by  DARPA grant No. W911NF--12--1--0034, AFOSR MURI grant No. FA9550-10-1-0569, and AFOSR award FA9550-11-1-0417.

\balancecolumns
\end{document}